\documentclass[prl,superscriptaddress,showpacs,twocolumn]{revtex4}
\usepackage{graphicx}
\def\beq{\begin{equation}}
\def\eeq{\end{equation}}
\def\be{\begin{equation}}
\def\ee{\end{equation}}

\def\iomn{i\omega_n}

\def\cG0{{\cal G}_0}
\def\cG{{\cal G}}

\def\spinup{\uparrow}
\def\spindown{\downarrow}

\def\Im{\mbox{Im}}

\def\nt{\widetilde{n}}
\def\Im{{\rm Im}}
\def\Jeff{\widetilde{J}}
\def\zeff{Z_{{\rm eff}}}

\begin{document}

\author{S. Biermann}
\affiliation{Centre de Physique Th{\'e}orique, {\'E}cole Polytechnique
91128 Palaiseau Cedex, France}
\author{L. de' Medici}
\affiliation{Centre de Physique Th{\'e}orique, {\'E}cole Polytechnique
91128 Palaiseau Cedex, France}
\affiliation{Laboratoire de Physique
des Solides, CNRS-UMR 8502, UPS B{\^a}t. 510, 91405 Orsay France}
\author{A. Georges}
\affiliation{Centre de Physique Th{\'e}orique, {\'E}cole Polytechnique
91128 Palaiseau Cedex, France}

\title{Non-Fermi Liquid Behavior and Double-Exchange Physics\\
in Orbital-Selective Mott Systems}



\begin{abstract}
We study a multi-band Hubbard model in its orbital selective
Mott phase, in which localized electrons in a narrow band coexist
with itinerant electrons in a wide band. The low-energy physics of
this phase is shown to be given by a generalized
double-exchange model. The high-temperature disordered phase thus 
differs from a Fermi
liquid, and displays a finite scattering rate of the conduction electrons
at the Fermi level, which depends continuously on the spin anisotropy.
\end{abstract}

\maketitle

The Mott phenomenon, localization of electrons by their
mutual interactions, is a key phenomenon in correlated electron
materials. Recently, there has been great interest in the possibility
of an ``orbital-selective Mott phase'' (OSMP), in which a Mott transition
occurs within a restricted subset of quasi-degenerate orbitals, so that
a narrow band of localised electrons coexists with a wider band of
itinerant electrons. Originally proposed~\cite{Anisimov_OSMT} for the ruthenate
compound Ca$_{x}$Sr$_{2-x}$RuO$_4$, this possibility has been intensively
discussed since then~\cite{Liebsch_OSMT_0,Koga_OSMT,demedici_OSMT_2005,Ferrero_OSMT,Arita_OSMT,Knecht_OSMT}.

In this letter, we clarify the nature of such an orbital-selective
Mott phase. We show that for finite Hund's coupling the orbital-selective
Mott transition is accompanied by a breakdown of Fermi liquid
theory, which can be understood from a mapping onto an
effective double-exchange model at low energy. 
In the high-temperature disordered phase,
the electrons in the conduction band acquire a
finite lifetime, due to the scattering on the localised spins.
This behavior is eventually cut-off by
long-range ordering at low temperature. Detailed calculations within
dynamical mean-field theory (DMFT) are performed in order to demonstrate
these properties.

We consider the two-orbital Hubbard model for a broad and a narrow
band, described by creation operators $c^\dagger_{i\sigma}$ and
$d^\dagger_{i\sigma}$ respectively, with hamiltonian:
\begin{eqnarray}
H
&=&-\sum_{ij\sigma} \left(t_{ij}
c^\dagger_{i \sigma}c_{j  \sigma}
+t^{d}_{ij} d^\dagger_{i \sigma}d_{j \sigma}
\right) +
\\
\nonumber
&+ & \sum_i(U\sum_{m=c,d}\nt^m_{i\spinup}\nt^m_{i\spindown}
+ U^{\prime} \sum_{\sigma\sigma'} \nt^c_{i\sigma}\nt^d_{i\sigma'}
- J\sum_{\sigma} \nt^c_{i\sigma}\nt^d_{i\sigma}
)
\\
\nonumber
&+&\alpha\, [H_{{\rm flip}} + H_{{\rm pair}}]
\nonumber
\label{eq:ham}
\end{eqnarray}
In this expression, $U$ and $U^\prime$ are Coulomb interaction parameters
and $J$ is the Hund's rule coupling. We have used the
notation $\nt^c_{i\sigma}=c^\dagger_{i\sigma} c_{i\sigma}-1/2$
(similarly, $\nt^d_{i\sigma}$) and written the hamiltonian in the
manifestly particle-hole symmetric case corresponding to two electrons
per site ($n^c=n^d=1$), on which we shall focus in the following.
The spin-flip and pair-hopping terms of the hamiltonian read:
$H_{{\rm flip}}=
-J\sum_i[c^\dagger_{i\spinup}c_{i\spindown}d^\dagger_{i\spindown}d_{i\spinup}+{\rm
h.c}]$ and $H_{{\rm pair}}=
-J\sum_i[c^\dagger_{i\spinup}c^\dagger_{i\spindown}d_{i\spinup}d_{i\spindown}
+ {\rm h.c}]$. A variable parameter $\alpha$ has been introduced in front of these
terms, for further use. For a cubic environment and unbroken spin symmetry, the
relations $U^\prime=U-2J$ and $\alpha=1$ hold.
This model has been the subject of several recent studies using
dynamical mean-field theory (or equivalently, in the
limit of infinite lattice connectivity).
For anisotropic interactions ($U^{\prime} < U$ and finite $J$)
an orbital selective Mott phase is found
even at moderate values of the ratio of the two bandwidths
\cite{Koga_OSMT}. As shown recently \cite{demedici_OSMT_2005,Ferrero_OSMT}, for large
differences in the bandwidths the OSMP can be induced
even in the case of $SU(4)$-symmetric interactions ($U=U^{\prime},J=0$).

\begin{figure}[h]
\centering
{\raisebox{0.0cm}{\resizebox{8.0cm}{!}
{\rotatebox{0}{\includegraphics{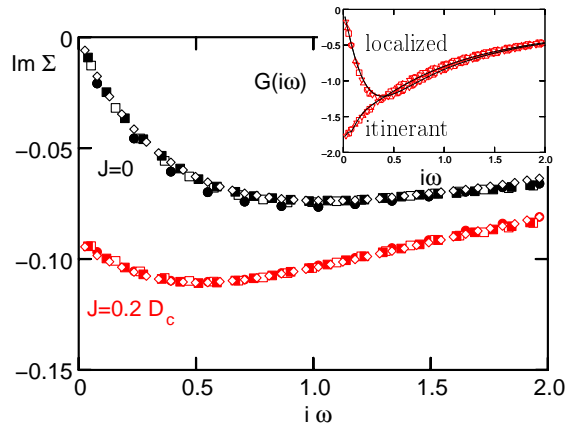}}}}}
\caption{
Broad band self-energy
$\Im\Sigma_c(\iomn)$ for different
temperatures ($\beta D_c = 40, 60, 80, 120$ correspond to
circles, empty and filled squares and diamonds respectively).
In the $J=0$ case (online: black)
$\Im\Sigma_c(i\omega)$ extrapolates linearly to zero
(Fermi liquid behavior), whereas for $J=0.2$ (online: red) a finite lifetime
is found.
Inset: $\Im G_d(\iomn)$ and $\Im G_c(\iomn)$
for U=0.8 D$_c$ and
J=0.2 D$_c$. QMC results (symbols) correspond
to finite temperatures ($\beta D_c = 40, 60, 80, 120$ given by
circles, squares, triangles up and down respectively),
while the ED data (lines) are at zero temperature.
}
\label{fig:ImSig}
\end{figure}

\begin{figure}[h]
\centering
{\raisebox{0.0cm}{\resizebox{7.0cm}{!}
{\rotatebox{0}{\includegraphics{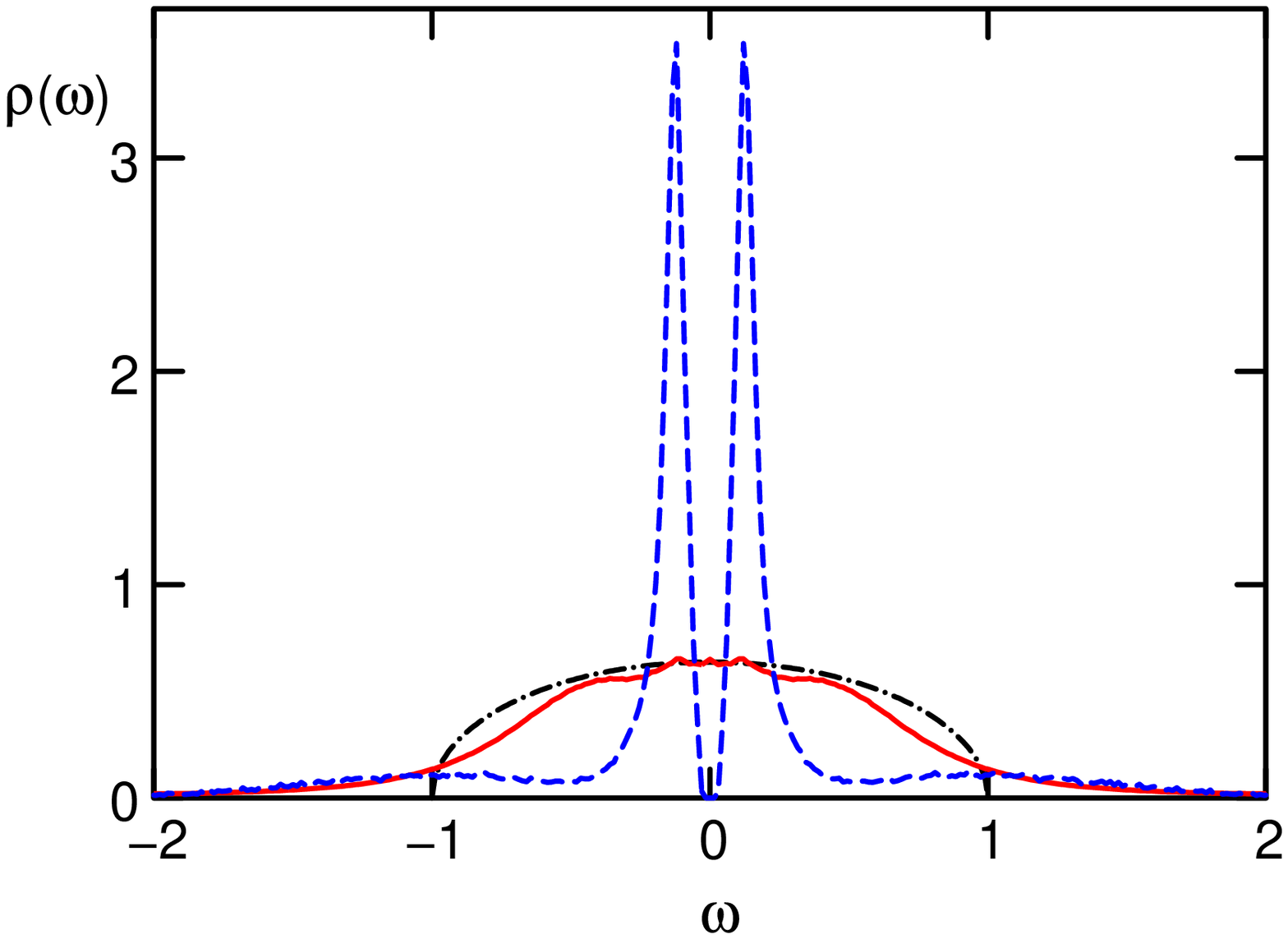}     
}}}}
{\raisebox{0.0cm}{\resizebox{7.0cm}{!}
{\rotatebox{0}{\includegraphics{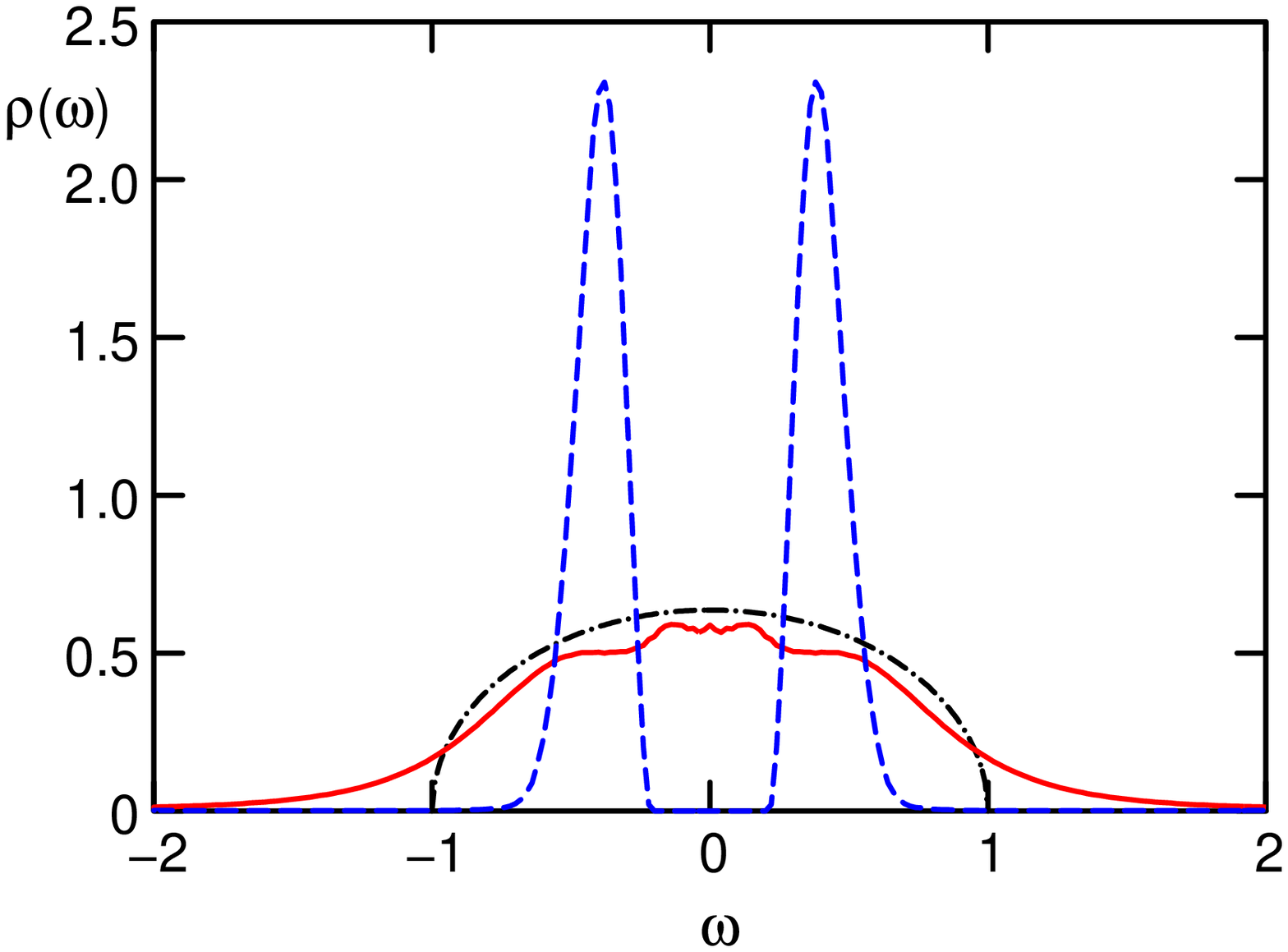}     
}}}}
\caption{Spectral functions for $J=0$ and $J=0.2$
($U=0.8D_c, \beta D_c=120$) calculated from QMC.
The narrow band $\rho_d(\omega)$ (dashed line) displays a gap or
pseudogap in both cases.
For $J=0$, $\rho_c(\omega)$ (solid line) coincides at $\omega=0$ with the
non-interacting density of states (dashed-dotted line), as a consequence of
Luttinger theorem, while this condition is clearly violated at
$J\neq 0$ ($\rho_c(0)<\rho_{0c}(0)$).
}
\label{fig:rho}
\end{figure}

First, we consider this model for $\alpha=0$, i.e with
density-density interactions only, and make a comparative
study of the nature of the paramagnetic OSMP in the absence
($J=0$) or in the presence ($J\neq 0$) of the Hund's coupling.
The model is solved using DMFT, for two semi-elliptical
densities of states (infinite connectivity Bethe lattice)
$\rho_{0c}(\epsilon)=2[1-(\epsilon/D_c)^2]^{1/2}/(\pi D_c)$
(similarly $\rho_{0d}$),
with a bandwidth ratio $D_d/D_c=0.1$. We use $D_c=1$ as our unit
of energy, choose $U=0.8$,
$U^{\prime} = U - 2 J$, and compare $J=0$ to $J=0.2$.
In both cases, the self-energy of the narrow band (not shown) diverges at low
frequency, while the self-energy of the broad band does not, indicating
that the narrow band is Mott-localized while the broad band is not (OSMP regime).
However, the low-frequency behavior of the self-energy is
very different in each case, as shown on Fig.~\ref{fig:ImSig} which
displays our DMFT results obtained with the QMC algorithm of Hirsch and Fye\cite{hirsch_fye}.
It is seen that, for $J=0$, the self-energy extrapolates to zero at
low frequency ($\Im\Sigma_c(\iomn)\sim (1-1/Z_c)\omega_n+\cdots$) in a
Fermi-liquid manner, while for $J\neq 0$, it extrapolates to a
finite value $\Im\Sigma_c(i0^+)\equiv\Gamma_c\neq 0$. Hence, a finite lifetime is
found at the Fermi level for $J\neq 0$, so that
no well-defined Fermi-liquid quasiparticles exist.
Correspondingly, this implies a violation of the
Luttinger sum-rule $\rho_c(0)=\rho_{c0}(0)$ for the spectral function
$\rho_c(\omega)\equiv-\Im G_c(\omega+i0^+)/\pi$.
The local Green's function being related
to the self-energy by:
$G_c(\iomn)=\int d\epsilon \rho_{c0}(\epsilon)/(\iomn-\Sigma_c(\iomn)-\epsilon)$,
a non-zero lifetime implies:
$\rho_c(0)=\int d\epsilon \rho_{c0}(\epsilon)
\Gamma_c/[\pi(\epsilon^2+\Gamma_c^2)]<\rho_{c0}(0)$.
On Fig.~\ref{fig:rho}, we display the spectral functions of the narrow and
broad bands, obtained by a maximum-entropy continuation of the QMC data.
Pinning at the Luttinger theorem value
$\rho_c(0)=\rho_{c0}(0)=2/(\pi D_c)$ holds for $J=0$ but not for
$J\neq 0$.
This is consistent with Fig.3 in \cite{Knecht_OSMT}.
Note that the spectral function of the (localised) narrow band has a sharp Mott
gap in the latter case, but displays rather a pseudogap in the former (as previously
discussed in \cite{demedici_OSMT_2005,Ferrero_OSMT}).
The Green's functions obtained
for $J=0.2$ using QMC and using an exact diagonalisation (ED) solution of the DMFT
equations (inset of Fig.~\ref{fig:ImSig})
are in perfect agreement.

In order to understand the origin of this finite-lifetime, non-Fermi liquid behavior, we
now derive an effective hamiltonian valid at low-energy in the orbital-selective regime. There, it
is expected to be legitimate to neglect entirely the hopping term of the narrow band
(which renormalizes to zero at low-energy). Furthermore, since we are interested
in low-energy physics, the states with a hole or a double-occupancy in the
narrow orbital can be eliminated.
Setting $t^d=0$ and $\nt^d_{\sigma}=\pm 1$ in (\ref{eq:ham}), a straightforward calculation
leads to the effective
hamiltonian:
\begin{eqnarray}
\nonumber
H_{\rm eff}&=&-\sum_{ij\sigma} t^c_{ij\sigma} c^\dagger_{i\sigma}c_{j \sigma}
+ U\sum_{i} \nt^c_{i\spinup}\nt^c_{\spindown} - \\
&-\,2J&\sum_{i} \left[S^z_{id}S^z_{ic}+\alpha(S^x_{id}S^x_{ic}+S^y_{id}S^y_{ic})\right]
\label{eq:hameff}
\end{eqnarray}
In this expression, $S^{x,y,z}_{id}$ are the components of the spin-1/2 operator acting
on the 2 local moment states $|\spinup\rangle_d,|\spindown\rangle_d$ at each site, while
$S^z_{ic}=1/2(c^\dagger_{i\spinup}c_{i\spinup}-c^\dagger_{i\spindown}c_{i\spindown})$,
$S^x_{ic}+iS^y_{ic}\equiv S^+_{ic}=c^\dagger_{i\spinup}c_{i\spindown}$ are the components
of the itinerant electron spin operator. Note that the inter-orbital Coulomb interaction
$U'$ entirely disappears from this effective hamiltonian (as well as the pair-hopping
term, which cannot act in the low-energy subspace of sites with single-occupancy of the d-band).
Hence the low-energy effective hamiltonian in the orbital-selective phase is
a {\it ferromagnetic Kondo lattice}, with an additional Hubbard interaction in the
(broad) itinerant band. The orbital-selective Mott localisation of a single
orbital component generates an effective
double-exchange model~\cite{zener_double_pr_1951}
in which the itinerant orbital interacts with the ``core-spin'' of the localised orbital through the
Hund's coupling.
This observation has two important consequences,
to be explored in more detail below: (i) it explains the observed non-Fermi liquid behavior,
and its dependence on the coupling to spin-flip terms $\alpha$ and (ii) it suggests that
ultimately the orbital-selective phase undergoes long-range ordering
as temperature is lowered.

\begin{figure}[h]
\centering
{\raisebox{0.0cm}{\resizebox{7.0cm}{!}
{\rotatebox{0}{
\includegraphics{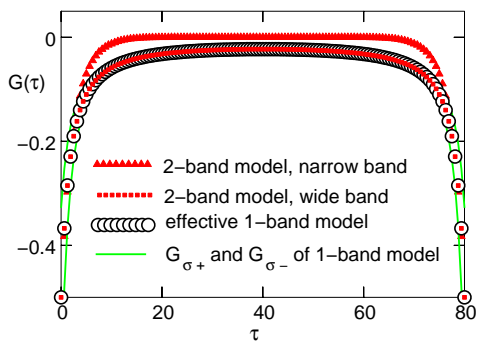}
}}}}
\caption{
Green's function in imaginary time for U=0.8D$_c$, J=0.2D$_c$,
$\beta=80$, computed from QMC. The red curves give the results
for the two-band model; the effective one-band model Green's
function (empty black circles) fit the wide band of the two-band
model perfectly. The green curves represent the auxiliary Green's
functions of the $S_z=\pm 1/2$ sectors of the effective 1-band
model.
}
\label{fig:rho_eff}
\end{figure}

When the interactions are of the density-density form ($\alpha=0$), the
itinerant orbital couples to the core-spin through Ising terms only.
The effective hamiltonian (\ref{eq:hameff}) then essentially reduces to a Falicov-Kimball model
for each spin species, in which the conduction electrons interact with a given
configuration of the Ising variables $S^d_{iz}$, which are conserved quantities.
In addition, the two spin components of the itinerant band are coupled
by the intra-orbital Hubbard interaction.
In order to solve this model, one must consider a given configuration of the Ising spins,
solve the one-band model of conduction electrons in that configuration, and average over all
Ising spin configurations with appropriate Boltzmann weight.
This process can be achieved within DMFT (infinite coordination) by mapping the problem onto
an effective single-site action:
\begin{eqnarray}
S_{{\rm eff}}^{\alpha=0} &=& -\int d\tau \int d\tau^{\prime}
\sum_{\sigma} \left(
c_{\sigma}^{\dagger} (\tau) \mathcal{G}_0^{-1}(\tau - \tau^{\prime})
c_{\sigma} (\tau^{\prime}) \right.
\nonumber
\\
&+&\int d\tau\left[U\nt_{\spinup}\nt_{\spindown}
- J S_{z} (n_{\spinup} - n_{\spindown}) \right]
\label{eq:seff}
\end{eqnarray}
subject to the self-consistency condition
$\mathcal{G}_0^{-1}(\iomn) = \iomn -t_c^2 G_c(\iomn)$.
Because $S^z$ is a conserved quantity, each sector $S^z=\pm 1/2$
can be considered independently. In each sector, the effective action
(\ref{eq:seff}) is that of an effective Anderson impurity model in a
local magnetic field $\pm J$. The itinerant electron Green's function is
obtained as: $G_c = 1/2[G_{\spinup +}+G_{\spinup -}]$,
where $G_{\sigma \pm}$ denotes the Green's function
$-\langle Tc^\dagger_{\sigma}c_{\sigma}\rangle_{\pm}$ in the effective
action (\ref{eq:seff}) with $S^z=\pm 1/2$, and the
obvious symmetry property $G_{\sigma +}=G_{\bar{\sigma} -}$ holds.
In writing these expressions, a paramagnetic state has been assumed and
particle-hole symmetry corresponding to half-filling has been used.
In Fig.~\ref{fig:rho_eff}, we compare the Green's function obtained by
solving this effective model to the full solution of the original model
in the OSMP.
The two results are indistinguishable for all
practical purposes, hence firmly
establishing the validity of our low-energy effective model in this phase.

For a given orientation of the core spin (say, $S^z=+1/2$), the
Green's functions of
the itinerant electrons for each spin component correspond to those of an impurity model
in a local magnetic field~\cite{laloux_mott_prb}, and hence read:
$G_{\spinup +}(\iomn)^{-1}=\mathcal{G}_0^{-1}+J/2-\Sigma_{\spinup +}(\iomn)$,
$G_{\spindown +}(\iomn)^{-1}=\mathcal{G}_0^{-1}-J/2-\Sigma_{\spindown +}(\iomn)$,
with the symmetry relation (at half-filling):
$\Sigma_{\spinup +}(\iomn)=-\Sigma_{\spindown +}(-\iomn)$.
In these expressions, $\Sigma_{\sigma +}$ is the self-energy originating from the
intra-orbital Hubbard interaction. In the itinerant (OSMP) phase for the
conduction electrons, these spin-resolved self-energies for a given core spin
orientation do have a Fermi-liquid form at low energy:
$\Sigma_{\spinup +}(\iomn)=\Sigma_0+(1-1/Z)\iomn+\cdots$, with $\Sigma_0$ a {\it real}
quantity~\cite{laloux_mott_prb}. $\Jeff=J-2\Sigma_0$ can be viewed as the effective exchange coupling renormalised
by the on-site Hubbard interaction.
Inserting these expressions into $G_c=1/2[G_{\spinup +}+G_{\spindown +}]$,
and taking into account the self-consistency condition $\mathcal{G}_0^{-1}=\iomn-t_c^2 G_c$
allows us to obtain the low-frequency form of the full self-energy $\Sigma_c$
in the form:
\begin{eqnarray}\nonumber
&\Sigma_c(\omega+i0^+)=-i\Gamma_c+(1-1/\zeff)\omega+\cdots\\
&\Gamma_c=\frac{\Jeff^2}{2(4t_c^2-\Jeff^2)^{1/2}}\,\,\,,\,\,\,
\zeff^{-1}=Z^{-1}\frac{(8t_c^2-\Jeff^2)(2t_c^2-\Jeff^2)}{(4t_c^2-\Jeff^2)^2}
\end{eqnarray}
Hence, a finite lifetime at zero energy emerges, due to the
scattering of the itinerant electrons onto the core spin associated
with the localised orbital.
Because the core $S^z$ is a conserved quantity in the absence of spin-flip
($\alpha=0$), the
physics is closely related to that of the Falicov-Kimball model, in which this
kind of non-Fermi liquid behavior was first emphasized by Si {\it et al.}~\cite{1992PhRvB..46.1261S},
and later by Furukawa~\cite{furukawa_double_dmft_jpsj_1994}
for double-exchange models in the context of manganites
(see also \cite{millis_double_prl_1995,millis_polaron2_prb_1996,
laad_srruo3,laad_vo2}).

\begin{figure}[h]
\centering
{\raisebox{0.0cm}{\resizebox{9.0cm}{!}
{\rotatebox{0}{
\includegraphics{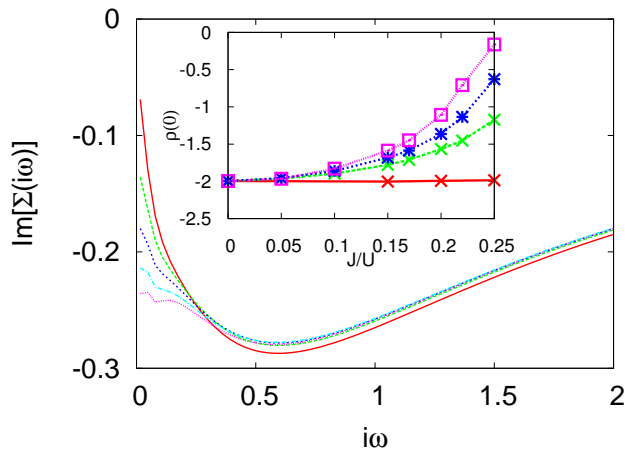}
}}}
}
\caption{
Low-energy part of $\Im\Sigma_c(i\omega)$, obtained by ED at $T=0$
for $U=1.2D_c, J=0.15 U$, and different
values of $\alpha$ ($0,0.3,0.5,0.7,1$ from bottom to top).
It extrapolates to zero at $\omega=0$ only in the
case of fully symmetric interactions ($\alpha=1$, red solid curve).
Inset: Zero-frequency extrapolation of the
Green's function
as a function of $J$ and for different values of $\alpha=1,0.7,0.5,0$
(from bottom to top).
In the case of a fully $SU(2)$ symmetric interaction vertex
($\alpha=1$) the Luttinger pinning condition $\Im G(0)=-2/D_c$
is fulfilled. As soon as this symmetry
is broken ($\alpha<1$), the pinning condition is lost.
}
\label{fig:anisotr}
\end{figure}

\begin{figure}[h]
\centering
{\raisebox{0.0cm}{\resizebox{9cm}{!}
{\rotatebox{0}{
\includegraphics{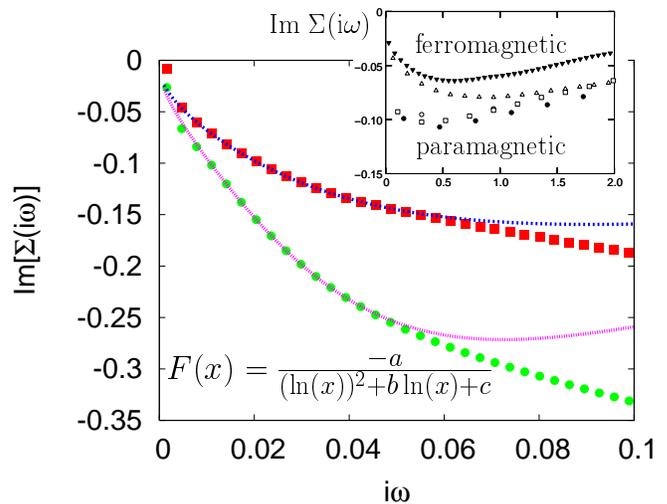}
}}}}
\caption{
Low-energy behavior of $\Im\Sigma_c(i\omega)$ obtained by ED
for $U=1.2D_c$ and $J=0.15 U$ (squares), $J=0.25 U$ (circles)
in the symmetric case
$\alpha$=1.
Solid lines are low-frequency fits of the data to
the function $F(\omega) = -a/[(\ln\omega)^2 + b\ln\omega+c]$.
{\it Inset}: Im$\Sigma_c(i\omega)$ for $U=0.8D_c,J=0.2D_c$, $\alpha=0$,
computed from QMC {\it without spin symmetrisation}.
Empty and solid circles, squares and triangles (up and down) correspond
to $\beta D_c=10, 20, 30, 40, 50, 120$. The last two temperatures
correspond to a ferromagnetic state.
}
\label{fig:mag}
\end{figure}

An obvious question is whether this finite lifetime is a robust feature
when spin-flip terms ($\alpha\neq 0$) are reintroduced.
In that case, we have to deal with a ferromagnetic Kondo
lattice model, with full $SU(2)$ symmetry at $\alpha=1$ and Ising anisotropy at $\alpha<1$.
Within DMFT, this model maps onto a ferromagnetic Kondo impurity problem, with a
self-consistent conduction electron bath. The renormalisation flow of the ferromagnetic
Kondo problem \cite{anderson_RGflow_1970} is such that the spin-flip component of the interaction is irrelevant as long
as $\alpha<1$, with a continuous family of infra-red fixed points corresponding
to X-ray edge physics. Hence, we expect a finite lifetime to survive as long as the
$SU(2)$ symmetry is broken (i.e for all $\alpha<1$). We have solved the DMFT equations for the
original 2-orbital model in the presence of spin-flip (and pair-hopping) terms for a range of values
of $\alpha$, in the orbital-selective phase. The results are displayed in Fig.\ref{fig:anisotr} and indeed
fully support this expectation, with a violation of Luttinger's sum rule gradually decreasing as
$\alpha$ is increased.
In the fully symmetric case $\alpha=1$, the lifetime
does extrapolate to zero (Fig.\ref{fig:anisotr}) as $\omega\rightarrow
0$,
but conventional Fermi liquid behavior is not recovered. Indeed,
single-impurity Kondo scaling suggests a singular behavior
$\Im\Sigma_c\sim -1/(\ln\omega)^2$, consistent with the fit of
our numerical results displayed in Fig.~\ref{fig:mag}.

The calculations described above are restricted to the
paramagnetic phase, spin symmetry being enforced on the Green's
functions. Clearly, at low temperatures, a phase with long-range
order is more favorable. The competition between ferromagnetism,
antiferromagnetism and phase separation has been studied in detail
for double exchange models relevant to manganites (see
e.g~\cite{furukawa_double_dmft_jpsj_1994,millis_polaron2_prb_1996,yunoki_phasesep_prl_1998,
dagotto_ferroKondo_prb_1998}). Fewer studies have taken into
account the effect of an on-site Hubbard interaction, which
increases the tendency to
ferromagnetism~\cite{dagotto_ferroKondo_prb_1998}. Such a detailed
study for the present model is beyond the scope of this paper. We
have however performed some calculations without spin- symmetrisation
and do find a ferromagnetic solution to be stable below a temperature
of order $D_c/50$. The inset of Fig.~\ref{fig:mag} demonstrates how a
ferromagnetic state, by suppressing spin disorder, restores a
scattering rate which decreases as temperature is lowered and as
the magnetic polarisation increases.

In conclusion, we have shown that the low-energy physics of
the orbital-selective Mott phase is given by a 
double exchange model (with an additional Hubbard repulsion).
As a result, in the high-temperature disordered phase 
a non-Fermi liquid state with a finite lifetime of
the conduction electrons is found.
At lower temperature, long-range spin ordering and possibly phase
separation sets in. By introducing a finite crystal field splitting,
it is clearly seen that the double- exchange physics and that of the OSMP
are continuously connected, the former corresponding to a large
crystal field splitting and a core spin being formed
in the lower crystal field level, while
the latter corresponds to a small crystal field and Mott localisation of the
narrow band.
Conversely, the OSMP is a useful way of simulating the double exchange model
with fully quantum spins using DMFT techniques.
Finally, it is tempting to speculate that a selective localisation
of electrons {\it in momentum space} may lead to short lifetimes over part of the
Fermi surface, as observed in the cuprates.

\acknowledgements
Calculations were performed at IDRIS Orsay (project number
051393). We thank A.~Millis for discussions.
While this work was being written, independent work by A.~Liebsch
suggested the OSMP to be a bad metal phase with non-Fermi liquid
behavior\cite{Liebsch_OSMT_3}.


\begin{thebibliography}{26}
\expandafter\ifx\csname natexlab\endcsname\relax\def\natexlab#1{#1}\fi
\expandafter\ifx\csname bibnamefont\endcsname\relax
  \def\bibnamefont#1{#1}\fi
\expandafter\ifx\csname bibfnamefont\endcsname\relax
  \def\bibfnamefont#1{#1}\fi
\expandafter\ifx\csname citenamefont\endcsname\relax
  \def\citenamefont#1{#1}\fi
\expandafter\ifx\csname url\endcsname\relax
  \def\url#1{\texttt{#1}}\fi
\expandafter\ifx\csname urlprefix\endcsname\relax\def\urlprefix{URL }\fi
\providecommand{\bibinfo}[2]{#2}
\providecommand{\eprint}[2][]{\url{#2}}
  
\bibitem[{\citenamefont{Anisimov et~al.}(2002)\citenamefont{Anisimov, Nekrasov,
  Kondakov, Rice, and Sigrist}}]{Anisimov_OSMT}
\bibinfo{author}{\bibfnamefont{V.}~\bibnamefont{Anisimov}},
  \bibinfo{author}{\bibfnamefont{I.}~\bibnamefont{Nekrasov}},
  \bibinfo{author}{\bibfnamefont{D.}~\bibnamefont{Kondakov}},
  \bibinfo{author}{\bibfnamefont{T.}~\bibnamefont{Rice}}, \bibnamefont{and}
  \bibinfo{author}{\bibfnamefont{M.}~\bibnamefont{Sigrist}},
  \bibinfo{journal}{Eur. Phys. J. B} \textbf{\bibinfo{volume}{25}},
  \bibinfo{pages}{191} (\bibinfo{year}{2002}).
  
\bibitem[{\citenamefont{Liebsch}(2003{\natexlab{a}})}]{Liebsch_OSMT_0}
\bibinfo{author}{\bibfnamefont{A.}~\bibnamefont{Liebsch}},
  \bibinfo{journal}{Europhysics Letters} \textbf{\bibinfo{volume}{63}},
  \bibinfo{pages}{97} (\bibinfo{year}{2003}{\natexlab{a}}),
  \bibinfo{journal}{Phys. Rev. Lett.} \textbf{\bibinfo{volume}{91}},
  \bibinfo{pages}{226401} (\bibinfo{year}{2003}{\natexlab{b}}),
  \bibinfo{journal}{Phys. Rev. B} \textbf{\bibinfo{volume}{70}},
  \bibinfo{pages}{165103} (\bibinfo{year}{2004}).
 
  
\bibitem[{\citenamefont{Koga et~al.}(2004{\natexlab{a}})\citenamefont{Koga,
  N.Kawakami, Rice, and Sigrist}}]{Koga_OSMT}
\bibinfo{author}{\bibfnamefont{A.}~\bibnamefont{Koga}},
  \bibinfo{author}{\bibnamefont{N.Kawakami}},
  \bibinfo{author}{\bibfnamefont{T.}~\bibnamefont{Rice}}, \bibnamefont{and}
  \bibinfo{author}{\bibfnamefont{M.}~\bibnamefont{Sigrist}},
  \bibinfo{journal}{Phys. Rev. Lett.} \textbf{\bibinfo{volume}{92}},
  \bibinfo{pages}{216402} (\bibinfo{year}{2004}{\natexlab{a}}),
  \bibinfo{note}{preprint
  cond-mat/0406457},
\bibinfo{note}{preprint cond-mat/0503651}.
  
\bibitem[{\citenamefont{de'~Medici et~al.}(2005)\citenamefont{de'~Medici,
  Georges, and Biermann}}]{demedici_OSMT_2005}
\bibinfo{author}{\bibfnamefont{L.}~\bibnamefont{de'~Medici}},
  \bibinfo{author}{\bibfnamefont{A.}~\bibnamefont{Georges}}, \bibnamefont{and}
  \bibinfo{author}{\bibfnamefont{S.}~\bibnamefont{Biermann}}
  (\bibinfo{year}{2005}), \eprint{condmat/0503764}.
  
\bibitem[{\citenamefont{Ferrero et~al.}(2005)\citenamefont{Ferrero, Becca,
  Fabrizio, and Capone}}]{Ferrero_OSMT}
\bibinfo{author}{\bibfnamefont{M.}~\bibnamefont{Ferrero}},
  \bibinfo{author}{\bibfnamefont{F.}~\bibnamefont{Becca}},
  \bibinfo{author}{\bibfnamefont{M.}~\bibnamefont{Fabrizio}}, \bibnamefont{and}
  \bibinfo{author}{\bibfnamefont{M.}~\bibnamefont{Capone}}
  (\bibinfo{year}{2005}), \bibinfo{note}{preprint cond-mat/0503759}.
  
\bibitem[{\citenamefont{Arita and Held}(2005)}]{Arita_OSMT}
\bibinfo{author}{\bibfnamefont{R.}~\bibnamefont{Arita}} \bibnamefont{and}
  \bibinfo{author}{\bibfnamefont{K.}~\bibnamefont{Held}}
  (\bibinfo{year}{2005}), \bibinfo{note}{preprint cond-mat/0504040}.
  
\bibitem[{\citenamefont{Knecht et~al.}(2005)\citenamefont{Knecht, Bl{\"u}mer, and
  van Dongen}}]{Knecht_OSMT}
\bibinfo{author}{\bibfnamefont{C.}~\bibnamefont{Knecht}},
  \bibinfo{author}{\bibfnamefont{N.}~\bibnamefont{Bl{\"u}mer}}, \bibnamefont{and}
  \bibinfo{author}{\bibfnamefont{P.}~\bibnamefont{van Dongen}}
  (\bibinfo{year}{2005}), \bibinfo{note}{preprint cond-mat/0505106}.
  
\bibitem[{\citenamefont{Hirsch and Fye}(1986)}]{hirsch_fye}
\bibinfo{author}{\bibfnamefont{J.~E.} \bibnamefont{Hirsch}} \bibnamefont{and}
  \bibinfo{author}{\bibfnamefont{R.~M.} \bibnamefont{Fye}},
  \bibinfo{journal}{Phys. Rev. Lett.} \textbf{\bibinfo{volume}{25}},
  \bibinfo{pages}{2521} (\bibinfo{year}{1986}).
  
\bibitem[{\citenamefont{Zener}(1951)}]{zener_double_pr_1951}
\bibinfo{author}{\bibfnamefont{C.}~\bibnamefont{Zener}},
  \bibinfo{journal}{Phys. Rev.} \textbf{\bibinfo{volume}{82}},
  \bibinfo{pages}{403} (\bibinfo{year}{1951}),
\bibinfo{author}{\bibfnamefont{P.~W.} \bibnamefont{Anderson}} \bibnamefont{and}
  \bibinfo{author}{\bibfnamefont{H.}~\bibnamefont{Hasegawa}},
  \bibinfo{journal}{Phys. Rev.} \textbf{\bibinfo{volume}{100}},
  \bibinfo{pages}{675} (\bibinfo{year}{1955}),
\bibinfo{author}{\bibfnamefont{P.~G.} \bibnamefont{{de Gennes}}},
  \bibinfo{journal}{Phys. Rev.} \textbf{\bibinfo{volume}{118}},
  \bibinfo{pages}{141} (\bibinfo{year}{1960}).
  
\bibitem[{\citenamefont{{Laloux} et~al.}(1994)\citenamefont{{Laloux},
  {Georges}, and {Krauth}}}]{laloux_mott_prb}
\bibinfo{author}{\bibfnamefont{L.}~\bibnamefont{{Laloux}}},
  \bibinfo{author}{\bibfnamefont{A.}~\bibnamefont{{Georges}}},
  \bibnamefont{and} \bibinfo{author}{\bibfnamefont{W.}~\bibnamefont{{Krauth}}},
  \bibinfo{journal}{Phys. Rev. B} \textbf{\bibinfo{volume}{50}},
  \bibinfo{pages}{3092} (\bibinfo{year}{1994}).
  
\bibitem[{\citenamefont{{Si} et~al.}(1992)\citenamefont{{Si}, {Kotliar}, and
  {Georges}}}]{1992PhRvB..46.1261S}
\bibinfo{author}{\bibfnamefont{Q.}~\bibnamefont{{Si}}},
  \bibinfo{author}{\bibfnamefont{G.}~\bibnamefont{{Kotliar}}},
  \bibnamefont{and}
  \bibinfo{author}{\bibfnamefont{A.}~\bibnamefont{{Georges}}},
  \bibinfo{journal}{Phys. Rev. B} \textbf{\bibinfo{volume}{46}},
  \bibinfo{pages}{R 1261} (\bibinfo{year}{1992}).
  
\bibitem[{\citenamefont{Furukawa}(1994)}]{furukawa_double_dmft_jpsj_1994}
\bibinfo{author}{\bibfnamefont{N.}~\bibnamefont{Furukawa}},
  \bibinfo{journal}{J. Phys. Soc. Jpn.} \textbf{\bibinfo{volume}{63}},
  \bibinfo{pages}{3214} (\bibinfo{year}{1994}).
  
\bibitem[{\citenamefont{{Millis} et~al.}(1995)\citenamefont{{Millis},
  {Littlewood}, and {Shraiman}}}]{millis_double_prl_1995}
\bibinfo{author}{\bibfnamefont{A.~J.} \bibnamefont{{Millis}}},
  \bibinfo{author}{\bibfnamefont{P.~B.} \bibnamefont{{Littlewood}}},
  \bibnamefont{and} \bibinfo{author}{\bibfnamefont{B.~I.}
  \bibnamefont{{Shraiman}}}, \bibinfo{journal}{Phys. Rev. Lett.}
  \textbf{\bibinfo{volume}{74}}, \bibinfo{pages}{5144} (\bibinfo{year}{1995}).
  
\bibitem[{\citenamefont{{Millis} et~al.}(1996)\citenamefont{{Millis},
  {Mueller}, and {Shraiman}}}]{millis_polaron2_prb_1996}
\bibinfo{author}{\bibfnamefont{A.~J.} \bibnamefont{{Millis}}},
  \bibinfo{author}{\bibfnamefont{R.}~\bibnamefont{{Mueller}}},
  \bibnamefont{and} \bibinfo{author}{\bibfnamefont{B.~I.}
  \bibnamefont{{Shraiman}}}, \bibinfo{journal}{Phys. Rev. B}
  \textbf{\bibinfo{volume}{54}}, \bibinfo{pages}{5405} (\bibinfo{year}{1996}).
  
\bibitem[{\citenamefont{Laad and M{\"u}ller-Hartmann}(2001)}]{laad_srruo3}
\bibinfo{author}{\bibfnamefont{M.~S.} \bibnamefont{Laad}} \bibnamefont{and}
  \bibinfo{author}{\bibfnamefont{E.}~\bibnamefont{M{\"u}ller-Hartmann}},
  \bibinfo{journal}{Phys. Rev. Lett.} \textbf{\bibinfo{volume}{87}},
  \bibinfo{pages}{246402} (\bibinfo{year}{2001}).
  
\bibitem[{\citenamefont{Laad et~al.}()\citenamefont{Laad, Craco, and
  M{\"u}ller-Hartmann}}]{laad_vo2}
\bibinfo{author}{\bibfnamefont{M.}~\bibnamefont{Laad}},
  \bibinfo{author}{\bibfnamefont{L.}~\bibnamefont{Craco}}, \bibnamefont{and}
  \bibinfo{author}{\bibfnamefont{E.}~\bibnamefont{M{\"u}ller-Hartmann}},
  \bibinfo{note}{preprints cond-mat/0409027 and 0505317}.
  
\bibitem[{\citenamefont{Anderson et~al.}(1970)\citenamefont{Anderson, Yuval,
  and Hamann}}]{anderson_RGflow_1970}
\bibinfo{author}{\bibfnamefont{P.~W.} \bibnamefont{Anderson}},
  \bibinfo{author}{\bibfnamefont{G.}~\bibnamefont{Yuval}}, \bibnamefont{and}
  \bibinfo{author}{\bibfnamefont{D.~R.} \bibnamefont{Hamann}},
  \bibinfo{journal}{Phys. Rev. B} \textbf{\bibinfo{volume}{1}},
  \bibinfo{pages}{4464} (\bibinfo{year}{1970}).
  
\bibitem[{\citenamefont{{Yunoki} et~al.}(1998)\citenamefont{{Yunoki}, {Hu},
  {Malvezzi}, {Moreo}, {Furukawa}, and {Dagotto}}}]{yunoki_phasesep_prl_1998}
\bibinfo{author}{\bibfnamefont{S.}~\bibnamefont{{Yunoki}}},
  \bibinfo{author}{\bibfnamefont{J.}~\bibnamefont{{Hu}}},
  \bibinfo{author}{\bibfnamefont{A.~L.} \bibnamefont{{Malvezzi}}},
  \bibinfo{author}{\bibfnamefont{A.}~\bibnamefont{{Moreo}}},
  \bibinfo{author}{\bibfnamefont{N.}~\bibnamefont{{Furukawa}}},
  \bibnamefont{and}
  \bibinfo{author}{\bibfnamefont{E.}~\bibnamefont{{Dagotto}}},
  \bibinfo{journal}{Phys. Rev. Lett.} \textbf{\bibinfo{volume}{80}},
  \bibinfo{pages}{845} (\bibinfo{year}{1998}).
  
\bibitem[{\citenamefont{{Dagotto} et~al.}(1998)\citenamefont{{Dagotto},
  {Yunoki}, {Malvezzi}, {Moreo}, {Hu}, {Capponi}, {Poilblanc}, and
  {Furukawa}}}]{dagotto_ferroKondo_prb_1998}
\bibinfo{author}{\bibfnamefont{E.}~\bibnamefont{{Dagotto}}},
  \bibinfo{author}{\bibfnamefont{S.}~\bibnamefont{{Yunoki}}},
  \bibinfo{author}{\bibfnamefont{A.~L.} \bibnamefont{{Malvezzi}}},
  \bibinfo{author}{\bibfnamefont{A.}~\bibnamefont{{Moreo}}},
  \bibinfo{author}{\bibfnamefont{J.}~\bibnamefont{{Hu}}},
  \bibinfo{author}{\bibfnamefont{S.}~\bibnamefont{{Capponi}}},
  \bibinfo{author}{\bibfnamefont{D.}~\bibnamefont{{Poilblanc}}},
  \bibnamefont{and}
  \bibinfo{author}{\bibfnamefont{N.}~\bibnamefont{{Furukawa}}},
  \bibinfo{journal}{Phys. Rev. B} \textbf{\bibinfo{volume}{58}},
  \bibinfo{pages}{6414} (\bibinfo{year}{1998}).
  
\bibitem[{\citenamefont{Liebsch}(2005)}]{Liebsch_OSMT_3}
\bibinfo{author}{\bibfnamefont{A.}~\bibnamefont{Liebsch}}
  (\bibinfo{year}{2005}), \bibinfo{note}{preprint cond-mat/0505393}.
  

 
\end{thebibliography}
\end{document}